\documentclass[12pt]{article}
\begin{document}

\title{\bf Energy-Momentum Distribution: A Crucial Problem
in General Relativity}

\author{M. Sharif \thanks{e-mail: msharif@math.pu.edu.pk} and Tasnim Fatima
\\ Department of Mathematics, University of the Punjab,\\ Quaid-e-Azam
Campus Lahore-54590, PAKISTAN.}

\date{}

\maketitle

\begin{abstract}
This paper is aimed to elaborate the problem of energy-momentum in
General Relativity. In this connection, we use the prescriptions
of Einstein, Landau-Lifshitz, Papapetrou and M\"{o}ller to compute
the energy-momentum densities for two exact solutions of Einstein
field equations. The spacetimes under consideration are the
non-null Einstein-Maxwell solutions and the singularity-free
cosmological model. The electromagnetic generalization of the
G\"{o}del solution and the G\"{o}del metric become special cases
of the non-null Einstein-Maxwell solutions. It turns out that
these prescriptions do not provide consistent results for any of
these spacetimes. These inconsistence results verify the
well-known proposal that the idea of localization does not follow
the lines of pseudo-tensorial construction but instead follows
from the energy-momentum tensor itself. These differences can also
be understood with the help of the Hamiltonian approach.
\end{abstract}


\date{}

\section{Introduction}

Energy-momentum is an important conserved quantity in any physical
theory whose definition has been under investigation for a long
time from the General Relativity (GR) viewpoint. The problem is to
find an expression which is physically meaningful. The point is
that the gravitational field can be made locally vanish and so one
is always able to find the frame in which the energy-momentum of
gravitational field is zero while in the other frames, it is not
true. Unfortunately, there is still no generally accepted
definition of energy-momentum for gravitational field. The problem
arises with the expression defining the gravitational field energy
part.

In the theory of GR, the energy-momentum conservation laws are
given by
\begin{equation}
T^b_{a;b}=0,\quad(a,b=0,1,2,3),
\end{equation}
where $T^{b}_a$ denotes the energy-momentum tensor. In order to
change the covariant divergence into an ordinary divergence so
that global energy-momentum conservation, including the
contribution from gravity, can be expressed in the usual manner as
in electromagnetism, Einstein formulated [1] the conservation law
in the following form
\begin{equation}
\frac{\partial}{\partial x^b}(\sqrt{-g}(T^b_a+t^b_a))=0.
\end{equation}
Here $t^b_a$ is not a tensor quantity and is called the
gravitational field pseudo-tensor. Schrodinger showed that the
pseudo-tensor can be made vanish outside the Schwarzschild radius
using a suitable choice of coordinates. There have been many
attempts in order to find a more suitable quantity for describing
the distribution of energy and momentum due to matter,
non-gravitational and gravitational fields. The proposed
quantities which actually fulfill the conservation law of matter
plus gravitational parts are called gravitational field
pseudo-tensors. The choice of the gravitational field
pseudo-tensor is not unique. Because of this, quite a few
definitions of these pseudo-tensors have been proposed.

In order to obtain a meaningful expression for energy, momentum
and angular momentum for a general relativistic system, Einstein
himself proposed an expression. After Einstein's energy-momentum
complex [2], many complexes have been found, for instance,
Landau-Lifshitz [3], Tolman [4], Papapetrou [5], M\"{o}ller [6,7],
Weinberg [8] and Bergman [9]. Some of these definitions are
coordinate dependent while others are not. Also, most of these
expression can not be used to define angular momentum.

The lack of a generally accepted definition of energy-momentum in
a curved spacetime has led to doubts regarding the idea of energy
localization. According to Misner et al. [10], energy is
localizable only for spherical systems. Cooperstock and Sarracino
[11] came up with the view that if energy is localizable for
spherical system, then it can be localized for any system. Bondi
[12] argued that a non-localizable form of energy is not allowed
in GR. After this, an alternative concept of energy, called
quasi-local energy, was developed. The use of quasi-local masses
to obtain energy-momentum in a curved spacetime do not restrict
one to use particular coordinate system.  A large number of
definitions of quasi-local masses have been proposed, those by
Penrose and many others [13-15]. Although these quasi-local masses
are conceptually very important, these definitions have serious
problems. Bergqvist [16] considered seven different definitions of
quasi-local masses and computed them for Reissner-Nordstrom and
Kerr spacetimes. He concluded that no two of the seven definitions
provide the same result. The seminal concept of quasi-local masses
of Penrose cannot be used to handle even the Kerr metric [17]. The
present quasi-local mass definitions still have inadequacies.

It is believed that different energy-momentum distribution would
be obtained from different energy-momentum complexes. Virbhadra
[18,19] revived the interest in this approach. He and his
co-workers [19-23] considered many asymptotically flat spacetimes
and showed that several energy-momentum complexes can give the
same result for a given spacetime. They also carried out
calculations in a few asymptotically non-flat spacetimes using
different energy-momentum complexes and found encouraging results.
Aguirregabiria et al. [24] proved that several energy-momentum
complexes can provide the same result for any Kerr-Schild class
metric. Chang et al. [25] showed that every energy-momentum
complex can be associated with a particular Hamiltonian boundary
term. Therefore, the energy-momentum complexes may also be
considered as quasi-local. Xulu [26,27] extended this
investigation and found that Melvin magnetic universe, Bianchi
type I universe provided the same energy distribution.

Virbhadra, Xulu and others [28] provided the hope that some
particular properties might give a basis to believe that some
pseudo-tensors of energy-momentum density had a special meaning.
Or equivalently that some coordinate exists which has a special
meaning. However, some examples of spacetimes have been explored
which do not support this viewpoint. In this regard, one of the
authors [29,30] considered the class of gravitational waves and
G\"{o}del universe, and used the two definitions of
energy-momentum. In a recent paper, the same author extended this
procedure to G\"{o}del-type metrics [31]. He concluded that both
the definitions do not provide consistent results for these
models. Ragab [32] also obtained similar results while dealing
with G\"{o}del-type metric using the prescriptions of M\"{o}ller
and Landau-Lifshitz. Contradictory results have also been obtained
by Owen [33] for a regular MMaS-class black hole.

According to the Hamiltonian approach, the various energy-momentum
expressions are each associated with distinct boundary conditions
[25,34]. It is found that using homogeneous boundary conditions,
the quasi-local energy vanishes for all Bianchi A but does not for
B models. Energy-momentum is associated with a closed surface
bounding a region. Energy can be identified as the value of the
Hamiltonian. The Hamiltonian for a finite region includes a
boundary term, which determines the quasi-local quantities and the
boundary conditions. In this paper, we are extending this work to
some more examples for the evaluation of energy-momentum density
components by using different energy-momentum complexes. We would
show that different prescriptions do not provide the same results
for a given spacetime which can be expected.

The paper is organized as follows. In section 2, we shall briefly
mention different prescriptions to evaluate energy-momentum
distribution. Sections 3 and 4 are devoted for the evaluation of
energy-momentum densities for the two particular spacetimes using
the prescriptions of Einstein, Landau-Lifshitz, Papapetrou and
M\"{o}ller. Finally, in the last section, we shall discuss and
summarize all the results obtained.

\section{Energy-Momentum Complexes}

In this section, we shall elaborate four different approaches to
evaluate the energy-momentum density components of different
spacetimes. We shall briefly describe the prescriptions of
Einstein, Landau-Lifshitz, Papapetrou and M\"{o}ller
energy-momentum complexes.

\subsection{Einstein Energy-Momentum Complex}

The energy-momentum complex of Einstein [2] is given by
\begin{equation}
\Theta^{b}_{a}=\frac{1}{16\pi}H^{bc}_{a,c},\quad
(a,b,...=0,1,2,3),
\end{equation}
where
\begin{equation}
H^{bc}_a=\frac{g_{ad}}{\sqrt{-g}}[-g(g^{bd}g^{ce}-g^{cd}g^{be})]_{,e}.
\end{equation}
It is to be noted that $H^{bc}_a$ is anti-symmetric in indices $b$
and $c$. $\Theta^{0}_{0}$ is the energy density,
$\Theta^{i}_{0}~(i=1,2,3)$ are the components of momentum density
and $\Theta^{0}_{i}$ are the energy current density components.
The Einstein energy-momentum satisfies the local conservation laws
\begin{equation}
\frac{\partial \Theta^{b}_{a}}{\partial x^b}=0.
\end{equation}
Einstein showed that the energy-momentum pseudo-complex
$\Theta^{b}_{a}$ provides satisfactory expression for the total
energy and momentum of closed system in the form of 3-dimensional
integral.

\subsection{Landau-Lifshitz Energy-Momentum Complex}

There were some drawbacks of Einstein energy-momentum complex. One
main drawback was that it was not symmetric in its indices. As a
result, this cannot be used to define conservation laws of angular
momentum. However, Landau-Lifshitz energy-momentum complex is
symmetric and they are able to develop a conserved angular
momentum complex in addition to that of energy-momentum. They
introduced a geodesic coordinate system at some particular point
in spacetime in which all the first derivatives of the metric
tensor vanish. The energy-momentum complex of Landau-Lifshitz [3]
is given by
\begin{equation}
L^{ab}= \frac{1}{16 \pi}\ell^{acbd}_{,cd},
\end{equation}
where
\begin{equation}
\ell^{acbd}= -g(g^{ab}g^{cd}-g^{ad}g^{cb}).
\end{equation}
$L^{00}$ represents the energy density of the whole system
including gravitation and $L^{oi}$ represent the components of the
total momentum density. $\ell^{abcd}$ has symmetries of the
Riemann curvature tensor. It is clear from Eq.(7) that $L^{ab}$ is
symmetric with respect to its indices. The energy-momentum complex
of Landau-Lifshitz satisfies the local conservation laws
\begin{equation}
\frac{\partial L^{ab}}{\partial x^b}=0.
\end{equation}

\subsection{Papapetrou Energy-Momentum Complex}

Papapetrou [5] energy-momentum complex is the least known among
the four definitions under discussion, as a result, it has been
re-discovered several times. The expression was found using the
generalized Belinfante method. The symmetric energy-momentum
complex of Papapetrou [5] is given as
\begin{equation}
\Omega^{ab}=\frac{1}{16\pi}N^{abcd}_{,cd},
\end{equation}
where
\begin{equation}
N^{abcd}=\sqrt{-g}(g^{ab}\eta^{cd}-g^{ac}\eta^{bd}
+g^{cd}\eta^{ab}-g^{bd}\eta^{ac}),
\end{equation}
and $\eta^{ab}$ is the Minkowski spacetime. The quantities
$N^{abcd}$ are symmetric in its first two indices $a$ and $b$. The
locally conserved quantities $\Omega^{ab}$ contain contribution
from the matter, non-gravitational and gravitational field. The
quantity $\Omega^{00}$ represents energy density and $\Omega^{0i}$
are the momentum density components. The energy-momentum complex
satisfies the local conservation laws
\begin{equation}
\frac{\partial \Omega^{ab}}{\partial x^b}=0.
\end{equation}

\subsection{M\"{o}ller Energy-Momentum Complex}

Although the Einstein energy-momentum complex provides useful
expression for the total energy-momentum of a closed system.
However, from the GR viewpoint, M\"{o}ller [7] argued that it is
unsatisfactory to transform a system into quasi-Cartesian
coordinates. M\"{o}ller tried to find out an expression of
energy-momentum which is independent of the choice of particular
coordinate system. His energy-momentum complex is given by
\begin{equation}
M^{b}_{a}=\frac{1}{8\pi}K^{bc}_{a,c},
\end{equation}
where
\begin{eqnarray}
K^{bc}_{a}=\sqrt{-g}(g_{ad,e}-g_{ae,d})g^{be}g^{cd}.
\end{eqnarray}
Here $K^{bc}_{a}$ is antisymmetric, $M^0_0$ is the energy density,
$M^i_0$ are the momentum density components and $M^0_i$ are the
energy current density components. The local conservation laws for
M\"{o}ller energy-momentum complex are the following
\begin{equation}
\frac{\partial M^{b}_{a}}{\partial x^b}=0.
\end{equation}
After a critical analysis of M\"{o}ller's result, Kovacs [35]
claimed that he found a mistake in M\"{o}ller's calculation.
However, Novotny [36] showed that M\"{o}ller [7] was right in
concluding that $p_a$ transforms like a four-vector under Lorentz
transformation. Lessner [37] showed that the problem is with the
interpretation of the result. He argued that energy-momentum
four-vector can transform according to Special Relativity only if
it is transformed to a reference system with the velocity constant
everywhere. He also concluded that the M\"{o}ller's
energy-momentum complex is a powerful expression of energy and
momentum in GR.

\section{Energy-Momentum Distribution in Non-Null
Einstein-Maxwell Solutions}

In this section, we calculate the energy-momentum densities for
the non-null Einstein-Maxwell solutions by using the four
different prescriptions given in the last section. Further, we
consider the two special cases of this solution and evaluate the
energy-momentum density components for these metrics. The non-null
Einstein-Maxwell solution contains five classes of non-null
electromagnetic field plus perfect fluid solutions which possesses
a metric symmetry not inherited by the electromagnetic field and
admits a homothetic vector field. Two of them contain electrovac
solutions as special cases, while the other three necessarily
contain fluid. This metric, representing the vacuum solution of
the Einstein field equations, is generalized by Kramer et al. [38]
and can be obtained by applying a complex invariance
transformation.

The line element of the non-null Einstein-Maxwell solutions [39]
is given by
\begin{equation}
ds^{2}= -(dt+Ad\phi)^2+F^2d\phi^2+e^{2K}(d\rho^{2}+dz^{2}),
\end{equation}
where $F=F(\rho)$, $A=A(\rho,z)$ and $K=K(\rho,z)$ are the
functions satisfying
\begin{eqnarray}
A_{,1}=FV_{,3},\quad A_{,3}=-FV_{,1},\nonumber\\
K_{,1}=-\frac{1}{4}a F(V^{2}_{,1}-V^{2}_{,3}),\nonumber\\
K_{,3}=-\frac{1}{2}aFV_{,1}V_{,3},\nonumber\\
V_{,11}+F_{,1}F^{-1}V_{,1}+V_{,33}=0.
\end{eqnarray}
In order to get meaningful results in the prescriptions of
Einstein, Landau-Lifshitz and Papapetrou one needs to have the
metric in Cartesian coordinates. For this reason, we transform the
metric in Cartesian coordinates by using the following
transformations
\begin{equation}
x=\rho\cos\phi,\quad y=\rho\sin\phi.
\end{equation}
The line element in $t,x,y,z$ coordinates becomes
\begin{eqnarray}
ds^{2}&=&dt^{2}+(A^{2}-F^{2})(\frac{xdy-ydx}{\rho^{2}})^{2}-e^{2K}
(\frac{xdx+ydy}{\rho})^{2}\nonumber\\
&-& e^{2K}dz^{2}+2Adt(\frac{xdy-ydx}{\rho^{2}}).
\end{eqnarray}

\subsection{Energy and Momentum in Einstein's Prescription}

In order to calculate the energy and momentum density components
for the non-null Einstein-Maxwell solutions, we need to compute
the components of $H^{bc}_{a}$. The required non-zero components
of $H^{bc}_{a}$ are
\begin{eqnarray}
H^{01}_{0}&=&-\frac{2x}{\rho^2}F_{\rho}-\frac{2xF}{\rho^2}K_{\rho}
+\frac{xF}{\rho^{3}}+\frac{xA}{\rho^{2}F}A_\rho+\frac{x}{\rho F}e^{2K},\\
H^{02}_{0}&=&-\frac{2y}{\rho^2}F_{\rho}-\frac{2yF
}{\rho^2}K_{\rho}
+\frac{yF}{\rho^{3}}+\frac{yA}{\rho^{2}F }A_\rho  +\frac{y}{\rho F }e^{2K},\\
H^{03}_{0}&=&\frac{A}{\rho F}A_{z}-\frac{2F}{\rho}K_z,\\
H^{12}_{0}&=&-H^{21}_{0}=\frac{A_{\rho}}{F},\\
H^{13}_{0}&=&-H^{31}_{0}=\frac{y}{\rho F}A_{z},\\
H^{23}_{0}&=&-H^{32}_{0}=-\frac{x}{\rho F}A_{z},\\
H^{01}_{1}&=&\frac{2Axy}{\rho^4}F_{\rho}-\frac{A^{2}xy}{\rho^4F}A_{\rho}
-\frac{Fxy}{\rho^{4}}A_{\rho},\\
H^{02}_{1}&=&\frac{2Ay^{2}}{\rho^4}F_{\rho}-\frac{A^{2}y^{2}}{\rho^4F}A_{\rho}
-\frac{Fy^{2}}{\rho^{4}}A_{\rho}-\frac{A}{\rho F}e^{2K},\\
H^{03}_{1}&=&-\frac{A^{2}y}{\rho^3 F}A_{z}-\frac{Fy}{\rho^{3}}A_{z},\\
H^{01}_{2}&=&-\frac{2Ax^{2}}{\rho^4}F_{\rho}+\frac{A^{2}x^{2}}{\rho^4F}A_{\rho}
+\frac{Fx^{2}}{\rho^{4}}A_{\rho}+\frac{A}{\rho F}e^{2K},\\
H^{02}_{2}&=&-\frac{2Axy}{\rho^4}F_{\rho}+\frac{A^{2}xy}{\rho^4F}A_{\rho}
+\frac{Fxy}{\rho^{4}}A_{\rho},\\
H^{03}_{2}&=&\frac{A^{2}x}{\rho^3F}A_{z}+\frac{Fx}{\rho^{3}}A_{z}.
\end{eqnarray}
Substituting Eqs.(19)-(30) in Eq.(3), we obtain the components of
energy and momentum density in the prescription of Einstein as
follows
\begin{eqnarray}
\Theta^0_{0}&=&\frac{1}{16 \pi \rho^{3} F^{2}}[F^{2}(-2 \rho^{2}
F_{\rho\rho}
-2 \rho^{2} F_{\rho} K_\rho-2\rho^{2}F K_{\rho\rho}+\rho F_{\rho}-F)\nonumber\\
&+&\rho^{2}FA^{2}_{\rho}-\rho^{2}AA_{\rho}F_{\rho}
+\rho^{2}Fe^{2K}+2\rho^{3}FK_{\rho}-\rho^{3}e^{2K}F_{\rho}\nonumber\\
&+&\rho^{2}AFA_{\rho\rho}+\rho^{2}F(A^{2}_{z}+AA_{zz}-2F^{2}K_{zz})],\\
\Theta^1_{0}&=&\frac{y}{16\pi \rho F^{2}}
(FA_{\rho\rho}-A_{\rho}F_{\rho}+FA_{zz}),\\
\Theta^2_{0}&=&-\frac{x}{16\pi \rho F^{2}}
(FA_{\rho\rho}-A_{\rho}F_{\rho}+FA_{zz}),\\
\Theta^0_{1}&=&\frac{y}{16\pi \rho^{4}F^{2}}
[F^{2}(-2AF_{\rho}+\rho A_{\rho}F_{\rho}
+2\rho AF_{\rho\rho}+FA_{\rho}-\rho FA_{\rho\rho})\nonumber\\
&+& \rho A^{2} A_{\rho}F_{\rho}+A^{2}FA_{\rho}-2\rho
AFA^{2}_{\rho}-\rho A^{2}FA_{\rho\rho}+\rho(\rho
AF_{\rho}+AF\nonumber\\&-&\rho FA_{\rho}-2\rho AF K_{\rho})e^{2K}
-\rho F(2AA^{2}_{z}+A^{2}A_{zz}+F^{2}A_{zz})],\\
\Theta^0_{2}&=&-\frac{x}{16\pi\rho^{4} F^{2}}
[F^{2}(-2AF_{\rho}+\rho A_{\rho}F_{\rho}
+2\rho AF_{\rho\rho}+FA_{\rho}-\rho FA_{\rho\rho})\nonumber\\
&+& \rho A^{2} A_{\rho}F_{\rho}+A^{2}FA_{\rho}-2\rho
AFA^{2}_{\rho}-\rho A^{2}FA_{\rho\rho}+\rho(\rho
AF_{\rho}+AF\nonumber\\&-&\rho FA_{\rho}-2\rho AF K_{\rho})e^{2K}
-\rho F(2AA^{2}_{z}+A^{2}A_{zz}+F^{2}A_{zz})],
\end{eqnarray}
and
\begin{equation}
\Theta^0_{3}=0=\Theta^3_{0}.
\end{equation}
If we choose the values of $A, F, K$ such that
\begin{equation}
A=\frac{m}{n}e^{n\rho},\quad F=e^{n\rho},\quad K=0,
\end{equation}
where $m,n$ are arbitrary constants, then the metric given by
Eq.(15) reduces to the electromagnetic generalization of the
G\"{o}del solution [40] and is given by
\begin{equation}
ds^{2}=
-(dt+\frac{m}{n}e^{n\rho}d\phi)^2+e^{2n\rho}d\phi^{2}+d\rho^{2}+dz^{2}.
\end{equation}
The corresponding energy-momentum density components turn out to
be
\begin{eqnarray}
\Theta^0_{0}&=&\frac{1}{16\pi\rho^{3}}[(\rho^{2}-n\rho^{3})
e^{-n\rho}+(n\rho-2n^{2}\rho^{2}
-1+m^{2}\rho^{2})e^{n\rho}],\\
\Theta^0_{1}&=&\frac{mye^{2n\rho}}{16\pi n^{2}\rho^{4}}
[2n^{3}\rho-n^{2}+n\rho e^{-2n\rho}
-2m^{2}n\rho+m^{2}],\\
\Theta^0_{2}&=&-\frac{mxe^{2n\rho}}{16\pi
n^{2}\rho^{4}}[2n^{3}\rho-n^{2}+n\rho e^{-2n\rho}
-2m^{2}n\rho+m^{2}].
\end{eqnarray}
The remaining momentum and energy current density components are
zero.

When we choose the values of the metric functions $A, F, K$ such
that
\begin{equation}
A=e^{ar},\quad F=\frac{e^{ar}}{\sqrt{2}},\quad K=0,
\end{equation}
where $a$ is an arbitrary constant, the original metric reduces to
\begin{equation}
ds^{2}=-(dt+e^{ar}d\phi)^2+\frac{e^{2ar}}{2}d\phi^{2}+dr^{2}+dz^{2}.
\end{equation}
This metric is known as G\"{o}del metric presented by K. G\"{o}del
in 1949 which represents one of the rotating spacetimes. When we
replace these values in Eqs.(31)-(35), we obtain the same results
as given in [30].

\subsection{Energy and Momentum in Landau-Lifshitz's Prescription}

The following non-vanishing components of $\ell^{acbd}$ are
required to find energy-momentum densities in this prescription
\begin{eqnarray}
\ell^{0101}&=&\frac{e^{2K}}{\rho^{4}}(A^{2}x^2
-F^{2}x^2-\rho^{2}y^{2}e^{2K}),\\
\ell^{0202}&=&\frac{e^{2K}}{\rho^{4}}(A^{2}y^2
-F^{2}y^2-\rho^{2}x^{2}e^{2K}),\\
\ell^{0102}&=&\frac{xye^{2K}}{\rho^{4}}(A^{2}
-F^{2}+\rho^{2}e^{2K}),\\
\ell^{0303}&=&\frac{e^{2K}}{\rho^{2}}(A^{2}-F^{2}),\\
\ell^{0112}&=&\frac{Ax}{\rho^2}e^{2K},\\
\ell^{0212}&=&\frac{Ay}{\rho^2}e^{2K},\\
\ell^{0313}&=&\frac{Ay}{\rho^2}e^{2K},\\
\ell^{0323}&=&-\frac{Ax}{\rho^2}e^{2K}.
\end{eqnarray}
Substituting these values in Eq.(6), we obtain the energy and
momentum density components as follows
\begin{eqnarray}
L^{00}&=&\frac{e^{2K}}{8\pi\rho^{4}}[A^{2}-F^{2}
-2\rho(AA_{\rho}-FF_{\rho})-2\rho
K_{\rho}(A^{2}-F^{2})\nonumber\\
&+&\rho^{2}K_{\rho\rho}(A^{2}-F^{2})
+2\rho^{2}K^{2}_{\rho}(A^{2}-F^{2})+4\rho^{2}K_{\rho}
(AA_{\rho}\nonumber\\&-&FF_{\rho}) +\rho^{2}(A^{2}_{\rho }
-F^{2}_{\rho })+\rho^{2}(AA_{\rho\rho}-FF_{\rho\rho})
+\rho^{2}\{AA_{zz}\nonumber\\&+&A^{2}_{z}
+4AA_{z}K_{z}+K_{zz}(A^{2}-F^{2})+2K^{2}_{z}(A^{2}-F^{2})\}],\\
L^{01}&=& \frac{ye^{2K}}{16\pi\rho^{3}}[-A_{\rho}+\rho
(A_{\rho\rho}+A_{zz})+4\rho
(A_{\rho}K_{\rho}+A_{z}K_{z})\nonumber\\&+&2\rho
A(K_{\rho\rho}+K_{zz})+4\rho A
(K^{2}_{\rho}+K^{2}_{z})-2AK_{\rho}],\\
L^{02}&=&-\frac{xe^{2K}}{16\pi\rho^{3}}[-A_{\rho}+\rho
(A_{\rho\rho}+A_{zz})+4\rho
(A_{\rho}K_{\rho}+A_{z}K_{z})\nonumber\\&+&2\rho
A(K_{\rho\rho}+K_{zz})+4\rho A
(K^{2}_{\rho}+K^{2}_{z})-2AK_{\rho}],\\
 L^{03}&=&0.
\end{eqnarray}
The energy and momentum density components for the metric given by
Eq.(38) are
\begin{eqnarray} L^{00}&=& \frac{e^{2n\rho}}{8\pi
n^{2}\rho^{4}}[m^{2}
-2m^{2}n\rho+2m^{2}n^{2}\rho^{2}-2n^{4}\rho^{2}+2n^{3}\rho-n^{2}],\\
L^{01}&=&\frac{mye^{n\rho}}{16\pi \rho^{3}}(n\rho-1),\\
L^{02}&=&-\frac{mxe^{n\rho}}{16\pi\rho^{3}}(n\rho-1),\\
L^{03}&=& 0.
\end{eqnarray}
The energy and momentum density components in the prescription of
Landau-Lifshitz for the G\"{o}del metric take the form
\begin{eqnarray}
L^{00}&=&\frac{e^{2ar}}{16\pi r^4}[1+2ar(ar-1)],\\
L^{01}&=&\frac{ay}{16\pi r^3}(ar-1)e^{ar},\\
L^{02}&=&-\frac{ax}{16\pi r^3}(ar-1)e^{ar},\\
L^{03}&=& 0.
\end{eqnarray}

\subsection{Energy and Momentum in Papapetrou's Prescription}

In this prescription, the required non-zero components of
$N^{abcd}$, given by Eq.(10), are
\begin{eqnarray}
N^{0011}&=&\frac{A^{2}}{\rho F}e^{2K}-\frac{F}{\rho}e^{2K}
-\frac{Fx^{2}}{\rho^{3}}-\frac{y^{2}}{\rho F}e^{2K},\\
N^{0022}&=&\frac{A^{2}}{\rho F}e^{2K}-\frac{F}{\rho}e^{2K}
-\frac{Fy^{2}}{\rho^{3}}-\frac{x^{2}}{\rho F}e^{2K},\\
N^{0033}&=&\frac{A^{2}}{F\rho}e^{2K}-\frac{F}{\rho}e^{2K}
-\frac{F}{\rho},\\
N^{0012}&=&\frac{xy}{\rho F}e^{2K}-\frac{Fxy}{\rho^{3}},\\
N^{0121}&=&\frac{Ax}{\rho F}e^{2K},\\
N^{0122}&=&\frac{Ay}{\rho F}e^{2K},\\
N^{0133}&=&\frac{Ay}{\rho F}e^{2K},\\
N^{0233}&=&-\frac{Ax}{\rho F}e^{2K}.
\end{eqnarray}
Making use of the Eqs.(64)-(71) in Eq.(9), we obtain energy and
momentum densities in Papapetrou's prescription
\begin{eqnarray}
\Omega^{00}&=&\frac{1}{16\pi\rho^{3}
F^{3}}[\{2A\rho^2F^2(A_{\rho\rho}+A_{zz})+2\rho^2F^2(A^2_\rho+A^2_z)\nonumber\\
&+&8A\rho^2F^2(A_\rho K_\rho+A_zK_z)-4\rho^2AFA_\rho F_\rho+
4\rho^2A^2F^2(K^2_\rho+K^2_z)\nonumber\\&+&2\rho^2A^2F^2(K_{\rho\rho}+K_{zz})
-4\rho^2A^2FF_\rho
K_\rho-\rho^2A^2FF_{\rho\rho}+2\rho^2A^2F^2_\rho\nonumber\\
&+&F^2(A^2-F^2)+2\rho F^2(F^2-A^2)K_\rho-\rho
A^2FF_\rho-\rho^2F^3F_{\rho\rho}\nonumber\\
&-&2\rho^2F^4(K_{\rho\rho}+K_{zz})-2\rho^2F^3F_\rho
K_\rho-4\rho^2F^4(K^2_\rho+K^2_z)-\rho^3FF_\rho\nonumber\\
&+&2\rho^3F^2K_\rho
+\rho^2F^2\}e^{2K}-\rho^2F^3F_{\rho\rho}],\\
\Omega^{01}&=&\frac{ye^{2K}}{16\pi \rho^{3}F^{3}}[\rho F^2A_\rho
-AF^2+\rho^2F^2(A_{\rho\rho}+A_{zz})\nonumber\\
&-&2\rho^2FA_\rho F_\rho+4\rho^2F^2(A_\rho
K_\rho+A_zK_z)\nonumber\\
&+&4\rho^2AF^2K^{2}_{\rho}-4\rho^2AFF_\rho K_\rho +2\rho^2AF^2
(K_{\rho\rho}+K_{zz})\nonumber\\
&-&\rho AFF_\rho-\rho^2AFF_{\rho\rho}
+2\rho^2AF^{2}_\rho+2\rho AF^2K_\rho],\\
\Omega^{02}&=&-\frac{xe^{2K}}{16\pi \rho^{3}F^{3}}[\rho F^2A_\rho
-AF^2+\rho^2F^2(A_{\rho\rho}+A_{zz})\nonumber\\
&-&2\rho^2FA_\rho F_\rho+4\rho^2F^2(A_\rho
K_\rho+A_zK_z)\nonumber\\
&+&4\rho^2AF^2K^{2}_{\rho}-4\rho^2AFF_\rho K_\rho +2\rho^2AF^2
(K_{\rho\rho}+K_{zz})\nonumber\\
&-&\rho AFF_\rho-\rho^2AFF_{\rho\rho}
+2\rho^2AF^{2}_\rho+2\rho AF^2K_\rho],\\
\Omega^{03}&=&0.
\end{eqnarray}
For the electromagnetic generalization of the G\"{o}del solution,
substituting the values of $A,~F,~K$ in the above expressions, we
obtain
\begin{eqnarray}
\Omega^{00}&=&\frac{e^{n\rho}}{16\pi
n^{2}\rho^{3}}[m^{2}-m^{2}n\rho+m^{2}n^{2}\rho^{2}-2n^{4}
\rho^{2}\nonumber\\&+&n^{3}\rho-n^{2}
+(n^{2}\rho^{2}-n^{3}\rho^{3})e^{-2n\rho}],\\
\Omega^{01}&=&-\frac{my}{16\pi n\rho^{3}},\\
\Omega^{02}&=&\frac{mx}{16\pi n\rho^{3}},\\
\Omega^{03}&=&0.
\end{eqnarray}
If we substitute the values of the metric functions given by
Eq.(42), we obtain the same energy-momentum density given in [30].

\subsection{Energy and Momentum in M\"{o}ller's Prescription}

Since the M\"{o}ller's prescription is not restricted to use the
Cartesian coordinates and hence the original metric given by
Eq.(15) can be used to find energy-momentum distribution. The
required non-vanishing components of $K^{bc}_a$ are
\begin{eqnarray}
K^{01}_0&=&\frac{A}{F}A_{\rho},\\
K^{03}_0&=&\frac{A}{F}A_{z},\\
K^{21}_0&=&-\frac{A_{\rho}}{F},\\
K^{23}_0&=&-\frac{A_{z}}{F},\\
K^{01}_2&=& FA_{\rho}+\frac{A^{2}}{F}A_{\rho}-2AF_{\rho},\\
K^{03}_2&=&FA_{z}+\frac{A^{2}}{F}A_{z}.
\end{eqnarray}
Using the above results in Eq.(12), we get
\begin{eqnarray}
M^0_0&=&\frac{1}{8\pi
F^{2}}[AF(A_{\rho\rho}+F_{zz})+F(A^{2}_{\rho}
+A^{2}_{z})-AA_{\rho}F_{\rho}],\\
M^2_0&=&\frac{1}{8\pi F^{2}}[A_{\rho}F_{\rho}-F(A_{\rho\rho}+A_{zz})],\\
M^0_2&=&\frac{1}{8\pi
F^{2}}[F^{3}(A_{\rho\rho}+A_{zz})+A^{2}F(A_{\rho\rho}
+A_{zz})\nonumber\\
&+&2AF(A^{2}_{\rho}+A^{2}_{z})-2AF^{2}F_{\rho\rho}
-(A^{2}+F^2)A_{\rho}F_{\rho}],
\end{eqnarray}
and
\begin{equation}
M^{0}_1=0= M^{0}_3= M^{1}_0= M^{3}_0.
\end{equation}
The corresponding components of the energy-momentum density
components for the metric (38) are
\begin{eqnarray}
M^0_0&=&\frac{m^2}{8\pi}e^{n\rho},\\
M^2_0&=&\frac{me^{2n\rho}}{4\pi n}(m^{2}-n^{2}).
\end{eqnarray}
The rest of the components are zero.

The energy and momentum densities for the G\"{o}del solution are
\begin{equation}
M^0_0= \frac{a^2e^{ar}}{4\sqrt{2}\pi},\\
\end{equation}
\begin{equation}
M^{0}_2= \frac{a^2e^{2ar}}{4\sqrt{2}\pi},
\end{equation}
\begin{equation}
M^{0}_1=0= M^{0}_3= M^i_0.
\end{equation}

\section{Energy-Momentum Distribution in Singularity-Free
Cosmological Model}

In this section, we extend the same procedure, applied in the
previous section, to another spacetime which is also cylindrically
symmetric. We consider a cosmological model representing perfect
fluid solution of EFEs which is non-separable in co-moving
coordinates and has non-singular scalar curvature invariants. This
corresponds to a cylindrical symmetric spacetime filled with an
isotropic radiation perfect fluid. This model is different from
the model investigated by Senovilla [40] in 1990. Also, it is
geodesically complete and globally hyperbolic. It fulfils the
energy, generic and causual conditions.

The line element for a spacetime that admits an abelian
two-dimensional orthogonal transitive group of isometries acting
on spacelike surfaces can be written in the form [41]
\begin{eqnarray}
ds^2=e^{2K}(-dt^2+dr^2)+\rho^2e^{2U}d\phi^2+e^{-2U}dz^2
\end{eqnarray}
where $K$ and $U$ are functions of $t$ and $r$. In Cartesian
coordinates, it becomes
\begin{eqnarray}
ds^{2}=e^{2K}[dt^2-(\frac{xdx+ydy}{r})^2]-\rho^2e^{2U}
(\frac{xdy-ydx}{r^2})^2-e^{-2U}dz^2.
\end{eqnarray}

\subsection{Einstein's Prescription}

The required components of $H^{bc}_{a}$ are the following
\begin{eqnarray}
H^{01}_{0}&=&-H^{10}_{0}=\frac{x\rho}{r^3}-\frac{2x}{r^2}\rho_r
+\frac{x}{\rho r}e^{2(K-U)},\\
H^{02}_{0}&=&-H^{20}_{0}=\frac{y\rho}{r^3}-\frac{2y}{r^2}\rho_r
+\frac{y}{\rho r}e^{2(K-U)},\\
H^{01}_{1}&=&-\frac{2x^2}{r^3}\rho_t-\frac{2x^2\rho}{r^{3}}(K_t-U_t),\\
H^{02}_{1}&=&H^{01}_{2}=-\frac{2xy}{r^3}\rho_t
+\frac{2xy\rho}{r^{3}}(K_t-U_t),\\
H^{02}_{2}&=&-\frac{2y^2}{r^3}\rho_t-\frac{2x^2\rho}{r^{3}}(K_t-U_t).
\end{eqnarray}
Substituting Eqs.(97)-(101) in Eq.(3), we obtain the components of
energy and momentum density
\begin{eqnarray}
\Theta^0_{0}&=&\frac{1}{16 \pi \rho^{2}
r^{3}}[r\rho^{2}\rho_r-\rho^3-2r^2\rho^2\rho_{rr}+\{\rho
r^2\nonumber\\
&+&2\rho r^3(K_r-U_r)-r^3\rho_r\}e^{2(K-U)}],\\
\Theta^1_{0}&=&\frac{x}{16\pi
\rho^{2}r^3}[2r\rho^2\rho_{rt}-\rho^2\rho_t +\{r^2\rho_t
-2\rho r^2(K_t-U_t)\}e^{2(K-U)}],\\
\Theta^2_{0}&=&\frac{y}{16\pi \rho^{2}r^3}[2r\rho^2\rho_{rt}
-\rho^2\rho_t+\{r^2\rho_t-2\rho r^2(K_t-U_t)\}e^{2(K-U)}],\\
\Theta^0_{1}&=&\frac{x}{8\pi r^{3}}[\rho(K_t-U_t)-r\rho_{tr}],\\
\Theta^0_{2}&=&\frac{y}{8\pi r^{3}}[\rho(K_t-U_t)-r\rho_{tr}],
\end{eqnarray}
and
\begin{equation}
\Theta^0_{3}=0=\Theta^3_{0}.
\end{equation}

\subsection{Landau-Lifshitz's Prescription}

The required non-vanishing components of $\ell^{acbd}$ are
\begin{eqnarray}
\ell^{0101}&=&-\frac{1}{r^{4}}(y^2r^2e^{2(K-U)}+x^2\rho^2),\\
\ell^{0202}&=&-\frac{1}{r^{4}}(x^2r^2e^{2(K-U)}+y^2\rho^2),\\
\ell^{0102}&=&\frac{xy}{r^{4}}(r^2e^{2(K-U)}-\rho^2),\\
\ell^{0110}&=&\frac{1}{r^{4}}(y^2r^2e^{2(K-U)}+x^2\rho^2),\\
\ell^{0210}&=&-\frac{xy}{r^{4}}(r^2e^{2(K-U)}-\rho^2),\\
\ell^{0220}&=&\frac{1}{r^{4}}(x^2r^2e^{2(K-U)}+y^2\rho^2).
\end{eqnarray}
Using the above results in Eq.(6), the energy and momentum density
components become
\begin{eqnarray}
L^{00}&=&\frac{1}{8\pi\rho^{4}}[r^{3}e^{2(K-U)}(K_r-U_r)
-\rho^{2}+2r\rho\rho_r-r^2\rho\rho_{\rho\rho}-r^2\rho^2_r],\\
L^{01}&=&
\frac{x}{8\pi\rho^{4}}[r\rho\rho_{rt}+r\rho_t\rho_r-\rho\rho_t
-r^2e^{2(K-U)}(K_r-U_r)],\\
L^{02}&=&
\frac{y}{8\pi\rho^{4}}[r\rho\rho_{rt}+r\rho_t\rho_r-\rho\rho_t
-r^2e^{2(K-U)}(K_r-U_r)],\\
 L^{03}&=&0.
\end{eqnarray}

\subsection{Papapetrou's Prescription}

We require the following non-vanishing components of $N^{abcd}$ to
find the energy-momentum density components in the prescription of
Papapetrou
\begin{eqnarray}
N^{0011}&=&-\frac{y^{2}}{r\rho}e^{2(K-U)}-\frac{x^2\rho}{r^3}-\frac{\rho}{r},\\
N^{0022}&=&-\frac{x^{2}}{r\rho}e^{2(K-U)}-\frac{y^2\rho}{r^3}-\frac{\rho}{r},\\
N^{0012}&=&\frac{xy}{r\rho}e^{2(K-U)}-\frac{xy\rho}{r^3},\\
N^{0101}&=&\frac{y^{2}}{r\rho}e^{2(K-U)}+\frac{x^2\rho}{r^3}+\frac{\rho}{r},\\
N^{0202}&=&\frac{x^{2}}{r\rho}e^{2(K-U)}+\frac{y^2\rho}{r^3}+\frac{\rho}{r},\\
N^{0102}&=&\frac{xy\rho}{r^3}-\frac{xy}{r\rho}e^{2(K-U)}.
\end{eqnarray}
Making use of the Eqs.(118)-(123) in Eq.(9), we obtain energy and
momentum densities as follows
\begin{eqnarray}
\Omega^{00}&=&\frac{1}{16\pi
r^3\rho^2}[r\rho^2\rho_r-r^2\rho^2\rho_{rr}-\rho^3\nonumber\\
&+&\{r^2\rho+2r^3\rho(K_r-U_r)-r^3\rho_r\}e^{2(K-U)}],\\
\Omega^{01}&=&\frac{x}{16\pi r^{3}\rho^{2}}[2r\rho^2
\rho_{tr}-\rho^2\rho_t+\{r^2\rho_t-r^2\rho(K_t-U_t)\}e^{2(K-U)}],\\
\Omega^{02}&=&\frac{y}{16\pi r^{3}\rho^{2}}[2r\rho^2
\rho_{tr}-\rho^2\rho_t+\{r^2\rho_t-r^2\rho(K_t-U_t)\}e^{2(K-U)}],\\
\Omega^{03}&=&0.
\end{eqnarray}

\subsection{M\"{o}ller's Prescription}

The required non-vanishing components of $K^{bc}_a$ are
\begin{eqnarray}
K^{01}_0&=&2\rho K_r,\\
K^{01}_1&=&2\rho K_t.
\end{eqnarray}
Using these values in Eq.(12), we get
\begin{eqnarray}
M^0_0&=&\frac{1}{4\pi}[\rho_rK_r+\rho K_{rr}],\\
M^1_0&=&-\frac{1}{4\pi }[\rho K_{rt}+\rho_tK_r],\\
M^0_1&=&\frac{1}{4\pi }[\rho K_{tr}+\rho_rK_t],
\end{eqnarray}
and
\begin{equation}
M^{0}_2=0= M^{0}_3= M^{2}_0= M^{3}_0.
\end{equation}

\section{Summary and Discussion}

The problem of energy-momentum localization has been a subject of
many researchers but still remains un-resolved. Numerous attempts
have been made to explore a quantity which describes the
distribution of energy-momentum due to matter, non-gravitational
and gravitational fields. Many energy-momentum complexes have been
found [2-9] and the problem associated with the energy-momentum
complexes leads to the doubts about the idea of energy
localization. This problem first appeared in electromagnetism
which turns out to be a serious matter in GR due to the
non-tensorial quantities. Many researchers considered different
energy-momentum complexes and obtained encouraging results.
Virbhadra et al. [18-23] explored several spacetimes for which
different energy-momentum complexes show a high degree of
consistency in giving the same and acceptable energy-momentum
distribution.

This paper continues the investigation of comparing various
distributions presented in the literature. It is devoted to
discuss the burning problem of energy-momentum in the frame work
of GR and four different energy-momentum complexes have been used
to find the energy-momentum distribution. These prescriptions turn
out to be a powerful tool to evaluate energy-momentum for various
physical systems. However, this tool is not proved to be the best
for some systems. Keeping this point in mind, we have applied the
prescriptions of Einstein, Landau-Lifshitz, Papapetrou and
M\"{o}ller to investigate energy-momentum distribution for various
spacetimes.

We have obtained energy-momentum densities for the non-null
Einstein-Maxwell solutions using the above prescriptions. This
solution reduces to the electromagnetic generalization of the
G\"{o}del solution and G\"{o}del metric for particular values of
the metric functions. We have extended the same procedure of
evaluating the energy-momentum distribution for these special
solutions and also for the singularity-free cosmological model.
The summary of the results (only non-zero quantities) can be given
in the form of tables in the following:

\vspace{1.0cm} \noindent {\bf {\small Table 1(a)}} {\bf Non-null
Einstein-Maxwell Solutions: Einstein's Prescription}

\vspace{0.1cm}

\begin{center}
\begin{tabular}{|l|l|}
\hline {\bf Energy-Momentum Densities} & {\bf Expression}
\\ \hline $\Theta^0_0$ & $
\begin{array}{c}
\frac{1}{16\pi\rho^{3}F^{2}}[F^{2}(-2\rho^{2}F_{\rho\rho}-2
\rho^{2}F_{\rho} K_\rho-2\rho^{2}FK_{\rho\rho}\\
+\rho F_{\rho}-F)+\rho^{2}FA^{2}_{\rho}-\rho^{2}AA_{\rho}F_{\rho}
+\rho^{2}Fe^{2K}\\
+2\rho^{3}FK_{\rho}-\rho^{3}e^{2K}F_{\rho}+\rho^{2}AFA_{\rho\rho}\\
+\rho^{2}F(A^{2}_{z}+AA_{zz}-2F^{2}K_{zz})]
\end{array}$
\\ \hline $\Theta^1_0$ & $
\begin{array}{c}
\frac{y}{16\pi \rho F^{2}}(FA_{\rho\rho}
-A_{\rho}F_{\rho}+FA_{zz})
\end{array}$
\\ \hline $\Theta^2_0$ & $\begin{array}{c}
-\frac{x}{16\pi \rho
F^{2}}(FA_{\rho\rho}-A_{\rho}F_{\rho}+FA_{zz})
\end{array}$
\\ \hline $\Theta^0_1$ & $\begin{array}{c}
\frac{y}{16\pi \rho^{4}F^{2}}[F^{2}(-2AF_{\rho}+\rho
A_{\rho}F_{\rho}+2\rho AF_{\rho\rho}\\
+FA_{\rho}-\rho FA_{\rho\rho})+\rho A^{2}
A_{\rho}F_{\rho}\\
+A^{2}FA_{\rho}-2\rho AFA^{2}_{\rho}-\rho
A^{2}FA_{\rho\rho}+\rho(\rho AF_{\rho}\\
+AF-\rho FA_{\rho}-2\rho AF K_{\rho})e^{2K}\\
-\rho F(2AA^{2}_{z}+A^{2}A_{zz}+F^{2}A_{zz})]
\end{array}$
\\ \hline $\Theta^0_2$ & $\begin{array}{c}
-\frac{x}{16\pi\rho^{4} F^{2}}[F^{2}(-2AF_{\rho}+\rho
A_{\rho}F_{\rho}+2\rho AF_{\rho\rho}\\
+FA_{\rho}-\rho FA_{\rho\rho})+\rho A^{2}
A_{\rho}F_{\rho}\\
+A^{2}FA_{\rho}-2\rho AFA^{2}_{\rho}-\rho
A^{2}FA_{\rho\rho}+\rho(\rho AF_{\rho}\\
+AF-\rho FA_{\rho}-2\rho AF K_{\rho})e^{2K}\\
-\rho F(2AA^{2}_{z}+A^{2}A_{zz}+F^{2}A_{zz})]
\end{array}$
\\ \hline
\end{tabular}
\end{center}

\newpage
\noindent
{\bf {\small Table 1(b)}} {\bf Non-null Einstein-Maxwell
Solutions: Landau-Lifshitz's Prescription}

\vspace{0.1cm}

\begin{center}
\begin{tabular}{|l|l|}
\hline {\bf Energy-Momentum Densities} & {\bf Expression}
\\ \hline $L^{00}$ & $
\begin{array}{c}
\frac{e^{2K}}{8\pi\rho^{4}}[(A^{2}-F^{2})-2\rho(AA_{\rho}-FF_{\rho})\\-2\rho
K_{\rho}(A^{2}-F^{2})+\rho^{2}K_{\rho\rho}(A^{2}-F^{2})\\
+2\rho^{2}K^{2}_{\rho}(A^{2}-F^{2})+4\rho^{2}K_{\rho}(AA_{\rho}\\
-FF_{\rho})+\rho^{2}(A^{2}_{\rho
}-F^{2}_{\rho})+\rho^{2}(AA_{\rho\rho}\\-FF_{\rho\rho})
+\rho^{2}\{AA_{zz}+A^{2}_{z}
+4AA_{z}K_{z}\\+(A^{2}-F^{2})K_{zz}+2(A^{2}-F^{2})K^{2}_{z}\}]
\end{array}$
\\ \hline $L^{01}$&$
\begin{array}{c}
\frac{ye^{2K}}{16\pi\rho^{3}} [-A_{\rho}+\rho
(A_{\rho\rho}+A_{zz})\\+4\rho(A_{\rho}K_{\rho}+A_{z}K_{z})+2\rho
A(K_{\rho\rho}\\+K_{zz})+4\rho A
(K^{2}_{\rho}+K^{2}_{z})-2AK_{\rho}]
\end{array}$
\\\hline $L^{02}$&$\begin{array}{c}
-\frac{xe^{2K}}{16\pi\rho^{3}} [-A_{\rho}+\rho
(A_{\rho\rho}+A_{zz})\\+4\rho(A_{\rho}K_{\rho}+A_{z}K_{z})+2\rho
A(K_{\rho\rho}\\+K_{zz})+4\rho A
(K^{2}_{\rho}+K^{2}_{z})-2AK_{\rho}]
\end{array}$
\\ \hline
\end{tabular}
\end{center}

\newpage
\noindent
{\bf {\small Table 1(c)}} {\bf Non-null Einstein-Maxwell
Solutions: Papapetrou's Prescription}

\vspace{0.1cm}

\begin{center}
\begin{tabular}{|l|l|}
\hline {\bf Energy-Momentum Densities} & {\bf Expression}
\\ \hline $\Omega^{00}$ & $
\begin{array}{c}
\frac{1}{16\pi\rho^{3} F^{3}}[\{2A\rho^2F^2(A_{\rho\rho}+A_{zz})\\
+2\rho^2F^2(A^2_\rho+A^2_z) + 8A\rho^2F^2(A_\rho
K_\rho+A_zK_z)\\-4\rho^2AFA_\rho F_\rho+
4\rho^2A^2F^2(K^2_\rho+K^2_z)\\+2\rho^2A^2F^2(K_{\rho\rho}+K_{zz})
-4\rho^2A^2FF_\rho
K_\rho\\-\rho^2A^2FF_{\rho\rho}+2\rho^2A^2F^2_\rho
+F^2(A^2-F^2)\\+2\rho F^2(F^2-A^2)K_\rho-\rho
A^2FF_\rho-\rho^2F^3F_{\rho\rho}\\
-2\rho^2F^4(K_{\rho\rho}+K_{zz})-2\rho^2F^3F_\rho
K_\rho\\-4\rho^2F^4 (K^2_\rho+K^2_z)-\rho^3FF_\rho
+2\rho^3F^2K_\rho\\ +\rho^2F^2\}e^{2K}-\rho^2F^3F_{\rho\rho}]
\end{array}$
\\ \hline $\Omega^{01}$&$
\begin{array}{c}
\frac{ye^{2K}}{16\pi \rho^{3}F^{3}} [\rho
F^2A_\rho-AF^2+\rho^2F^2(A_{\rho\rho}+A_{zz})\\-2\rho^2FA_\rho
F_\rho+4\rho^2F^2(A_\rho K_\rho\\+A_zK_z)
+4\rho^2AF^2K^{2}_{\rho}-4\rho^2AFF_\rho K_\rho\\+2\rho^2AF^2
(K_{\rho\rho}+K_{zz})-\rho AFF_\rho\\-\rho^2AFF_{\rho\rho}
+2\rho^2AF^{2}_\rho+2\rho AF^2K_\rho]
\end{array}$
\\\hline $\Omega^{02}$&$\begin{array}{c}
-\frac{xe^{2K}}{16\pi \rho^{3}F^{3}}[\rho
F^2A_\rho-AF^2+\rho^2F^2(A_{\rho\rho}+A_{zz})\\-2\rho^2FA_\rho
F_\rho+4\rho^2F^2(A_\rho K_\rho\\+A_zK_z)
+4\rho^2AF^2K^{2}_{\rho}-4\rho^2AFF_\rho K_\rho\\+2\rho^2AF^2
(K_{\rho\rho}+K_{zz})-\rho AFF_\rho\\-\rho^2AFF_{\rho\rho}
+2\rho^2AF^{2}_\rho+2\rho AF^2K_\rho]
\end{array}$
\\ \hline
\end{tabular}
\end{center}

\vspace{0.5cm}

\noindent
{\bf {\small Table 1(d)}} {\bf Non-null Einstein-Maxwell
Solutions: M\"{o}ller's Prescription}

\vspace{0.1cm}

\begin{center}
\begin{tabular}{|l|l|}
\hline {\bf Energy-Momentum Densities} & {\bf Expression}
\\ \hline $M^0_0$ & $
\begin{array}{c}
\frac{1}{8\pi F^{2}}[AF(A_{\rho\rho}+F_{zz})+F(A^{2}_{\rho}
+A^{2}_{z})-AA_{\rho}F_{\rho}]
\end{array}$
\\ \hline $M^2_0$ & $
\begin{array}{c}
\frac{1}{8\pi F^{2}}[A_{\rho}F_{\rho}-F(A_{\rho\rho}+A_{zz})]
\end{array}$
\\ \hline $M^0_2$ & $\begin{array}{c}
\frac{1}{8\pi
F^{2}}[F^{3}(A_{\rho\rho}+A_{zz})+A^{2}F(A_{\rho\rho}
+A_{zz})\\-2AF^{2}F_{\rho\rho}+2AF(A^{2}_{\rho}+A^{2}_{z})\\
-(A^{2}+F^2)A_{\rho}F_{\rho}]
\end{array}$
\\ \hline
\end{tabular}
\end{center}

\newpage

\noindent
{\bf {\small Table 2(a)}} {\bf Electromagnetic
Generalization of the G\"{o}del solutions: Einstein's
Prescription}

\vspace{0.1cm}

\begin{center}
\begin{tabular}{|l|l|}
\hline {\bf Energy-Momentum Densities} & {\bf Expression}
\\ \hline $\Theta^0_0$ & $
\begin{array}{c}
\frac{1}{16\pi\rho^{3}}[(\rho^{2}-n\rho^{3})
e^{-n\rho}\\+(n\rho-2n^{2}\rho^{2} -1+m^{2}\rho^{2})e^{n\rho}]
\end{array}$
\\ \hline $\Theta^0_1$ & $\begin{array}{c}
\frac{mye^{2n\rho}}{16\pi n^{2}\rho^{4}}[2n^{3}\rho-n^{2}+n\rho
e^{-2n\rho} -2m^{2}n\rho+m^{2}]
\end{array}$
\\ \hline $\Theta^0_2$ & $\begin{array}{c}
-\frac{mxe^{2n\rho}}{16\pi n^{2}\rho^{4}}[2n^{3}\rho-n^{2}+n\rho
e^{-2n\rho} -2m^{2}n\rho+m^{2}]
\end{array}$
\\ \hline
\end{tabular}
\end{center}

\vspace{0.5cm}

\noindent
{\bf {\small Table 2(b)}} {\bf Electromagnetic
Generalization of the G\"{o}del solutions: Landau-Lifshitz's
Prescription}

\vspace{0.1cm}

\begin{center}
\begin{tabular}{|l|l|}
\hline {\bf Energy-Momentum Densities} & {\bf Expression}
\\ \hline $L^{00}$ & $
\begin{array}{c}
\frac{e^{2n\rho}}{8\pi n^{2}\rho^{4}}[m^{2}
-2m^{2}n\rho\\+2m^{2}n^{2}\rho^{2}-2n^{4}\rho^{2}+2n^{3}\rho-n^{2}]
\end{array}$
\\ \hline $L^{01}$&$
\begin{array}{c}
\frac{mye^{n\rho}}{16\pi \rho^{3}}(n\rho-1)
\end{array}$
\\\hline $L^{02}$&$\begin{array}{c}
-\frac{mxe^{n\rho}}{16\pi\rho^{3}}(n\rho-1)
\end{array}$
\\ \hline
\end{tabular}
\end{center}

\vspace{0.5cm}

\noindent
{\bf {\small Table 2(c)}} {\bf Electromagnetic
Generalization of the G\"{o}del solutions: Papapetrou's
Prescription}

\vspace{0.1cm}

\begin{center}
\begin{tabular}{|l|l|}
\hline {\bf Energy-Momentum Densities} & {\bf Expression}
\\ \hline $\Omega^{00}$ & $
\begin{array}{c}
\frac{e^{n\rho}}{16\pi
n^{2}\rho^{3}}[m^{2}-m^{2}n\rho+m^{2}n^{2}\rho^{2}-2n^{4}
\rho^{2}\\+n^{3}\rho-n^{2}
+(n^{2}\rho^{2}-n^{3}\rho^{3})e^{-2n\rho}]
\end{array}$
\\ \hline $\Omega^{01}$&$
\begin{array}{c}
-\frac{my}{16\pi n\rho^{3}}
\end{array}$
\\\hline $\Omega^{02}$&$\begin{array}{c}
\frac{mx}{16\pi n\rho^{3}}
\end{array}$
\\ \hline
\end{tabular}
\end{center}

\vspace{0.5cm}

\noindent
{\bf {\small Table 2(d)}} {\bf Electromagnetic
Generalization of the G\"{o}del solutions: M\"{o}ller's
Prescription}

\vspace{0.1cm}

\begin{center}
\begin{tabular}{|l|l|}
\hline {\bf Energy-Momentum Densities} & {\bf Expression}
\\ \hline $M^0_0$ & $
\begin{array}{c}
\frac{m^2}{8\pi}e^{n\rho}
\end{array}$
\\ \hline $M^2_0$ & $
\begin{array}{c}
\frac{me^{2n\rho}}{4\pi n}(m^{2}-n^{2})
\end{array}$
\\ \hline
\end{tabular}
\end{center}

\vspace{0.2cm}

\noindent
{\bf {\small Table 3(a)}} {\bf G\"{o}del Metric:
Landau-Lifshitz's Prescription}

\vspace{0.1cm}

\begin{center}
\begin{tabular}{|l|l|}
\hline {\bf Energy-Momentum Densities} & {\bf Expression}
\\ \hline $L^{00}$ & $
\begin{array}{c}
\frac{e^{2ar}}{16\pi r^4}[1+2ar(1-ar)]
\end{array}$
\\ \hline $L^{01}$&$
\begin{array}{c}
\frac{ay}{16\pi r^3}(ar-1)e^{ar}
\end{array}$
\\\hline $L^{02}$&$\begin{array}{c}
-\frac{ax}{16\pi r^3}(ar-1)e^{ar}
\end{array}$
\\ \hline
\end{tabular}
\end{center}

\vspace{0.5cm}

\noindent
{\bf {\small Table 3(b)}} {\bf G\"{o}del Metric:
M\"{o}ller's Prescription}

\vspace{0.1cm}

\begin{center}
\begin{tabular}{|l|l|}
\hline {\bf Energy-Momentum Densities} & {\bf Expression}
\\ \hline $M^0_0$ & $
\begin{array}{c}
\frac{a^2e^{ar}}{4\sqrt{2}\pi}
\end{array}$
\\ \hline $M^0_2$ & $
\begin{array}{c}
\frac{a^2e^{2ar}}{4\sqrt{2}\pi}
\end{array}$
\\ \hline
\end{tabular}
\end{center}

\vspace{0.5cm}

\noindent
{\bf {\small Table 4(a)}} {\bf Singularity-Free
Cosmological Model: Einstein's Prescription}

\vspace{0.1cm}

\begin{center}
\begin{tabular}{|l|l|}
\hline {\bf Energy-Momentum Densities} & {\bf Expression}
\\ \hline $\Theta^0_0$ & $
\begin{array}{c}
\frac{1}{16 \pi \rho^{2}
r^{3}}[r\rho^{2}\rho_r-\rho^3-2r^2\rho^2\rho_{rr}\\+\{\rho r^2
+2\rho r^3(K_r-U_r)-r^3\rho_r\}e^{2(K-U)}]
\end{array}$
\\ \hline $\Theta^1_0$ & $
\begin{array}{c}
\frac{x}{16\pi \rho^{2}r^3}[2r\rho^2\rho_{rt}-\rho^2\rho_t\\
+\{r^2\rho_t -2\rho r^2(K_t-U_t)\}e^{2(K-U)}]
\end{array}$
\\ \hline $\Theta^2_0$ & $\begin{array}{c}
\frac{y}{16\pi \rho^{2}r^3}[2r\rho^2\rho_{rt}
-\rho^2\rho_t\\+\{r^2\rho_t-2\rho r^2(K_t-U_t)\}e^{2(K-U)}]
\end{array}$
\\ \hline $\Theta^0_1$ & $\begin{array}{c}
\frac{x}{8\pi r^{3}}[\rho(K_t-U_t)-r\rho_{tr}]
\end{array}$
\\ \hline $\Theta^0_2$ & $\begin{array}{c}
\frac{y}{8\pi r^{3}}[\rho(K_t-U_t)-r\rho_{tr}]
\end{array}$
\\ \hline
\end{tabular}
\end{center}

\vspace{0.5cm}
\newpage
\noindent
{\bf {\small Table 4(b)}} {\bf Singularity-Free
Cosmological Model: Landau-Lifshitz's Prescription}

\vspace{0.1cm}

\begin{center}
\begin{tabular}{|l|l|}
\hline {\bf Energy-Momentum Densities} & {\bf Expression}
\\ \hline $L^{00}$ & $
\begin{array}{c}
\frac{1}{8\pi\rho^{4}}[r^{3}e^{2(K-U)}(K_r-U_r)-\rho^{2}+2r\rho\rho_r\\
-r^2\rho\rho_{\rho\rho}-r^2\rho^2_r]
\end{array}$
\\ \hline $L^{01}$&$
\begin{array}{c}
\frac{x}{8\pi\rho^{4}}[r\rho\rho_{rt}+r\rho_t\rho_r\\
-\rho\rho_t-r^2e^{2(K-U)}(K_r-U_r)]
\end{array}$
\\\hline $L^{02}$&$\begin{array}{c}
\frac{y}{8\pi\rho^{4}}[r\rho\rho_{rt}+r\rho_t\rho_r\\-\rho\rho_t
-r^2e^{2(K-U)}(K_r-U_r)]
\end{array}$
\\ \hline
\end{tabular}
\end{center}

\vspace{0.5cm}

\noindent
{\bf {\small Table 4(c)}} {\bf Singularity-Free
Cosmological Model: Papapetrou's Prescription}

\vspace{0.1cm}

\begin{center}
\begin{tabular}{|l|l|}
\hline {\bf Energy-Momentum Densities} & {\bf Expression}
\\ \hline $\Omega^{00}$ & $
\begin{array}{c}
\frac{1}{16\pi
r^3\rho^2}[r\rho^2\rho_r-r^2\rho^2\rho_{rr}-\rho^3\\
+\{r^2\rho+2r^3\rho(K_r-U_r)-r^3\rho_r\}e^{2(K-U)}]
\end{array}$
\\ \hline $\Omega^{01}$&$
\begin{array}{c}
\frac{x}{16\pi r^{3}\rho^{2}}[2r\rho^2
\rho_{tr}-\rho^2\rho_t\\
+\{r^2\rho_t-r^2\rho(K_t-U_t)\}e^{2(K-U)}]
\end{array}$
\\\hline $\Omega^{02}$&$\begin{array}{c}
\frac{y}{16\pi r^{3}\rho^{2}}[2r\rho^2
\rho_{tr}-\rho^2\rho_t\\
+\{r^2\rho_t-r^2\rho(K_t-U_t)\}e^{2(K-U)}]
\end{array}$
\\ \hline
\end{tabular}
\end{center}

\vspace{0.5cm}

\noindent
{\bf {\small Table 4(d)}} {\bf Singularity-Free
Cosmological Model: M\"{o}ller's Prescription}

\vspace{0.1cm}

\begin{center}
\begin{tabular}{|l|l|}
\hline {\bf Energy-Momentum Densities} & {\bf Expression}
\\ \hline $M^0_0$ & $
\begin{array}{c}
\frac{1}{4\pi}[\rho_rK_r+\rho K_{rr}]
\end{array}$
\\ \hline $M^1_0$ & $
\begin{array}{c}
-\frac{1}{4\pi }[\rho K_{rt}+\rho_tK_r]
\end{array}$
\\ \hline $M^0_1$ & $\begin{array}{c}
\frac{1}{4\pi }[\rho K_{tr}+\rho_rK_t]
\end{array}$
\\ \hline
\end{tabular}
\end{center}

\vspace{0.2cm}

From these results, it can be seen that the energy-momentum
density components turn out to be finite and well-defined in the
above mentioned prescriptions. The four prescriptions of the
energy-momentum complexes do not provide the same results for any
of these spacetimes. The energy-momentum densities for the
non-null Einstein-Maxwell solutions reduce to the energy-momentum
densities for the electromagnetic generalization of the G\"{o}del
solution and G\"{o}del metric for particular values of the metric
functions. We have also applied the same procedure to the
singularity-free cosmological model which also gives different
results in each prescriptions.

It is worth mentioning that the results of energy-momentum
distribution for different spaceimes are not surprising rather
they justify that different energy-momentum complexes, which are
pseudo-tensors, are not covariant objects. This is in accordance
with the equivalence principle [10] which implies that the
gravitational field cannot be detected at a point. These examples
indicate that the idea of localization does not follow the lines
of pseudo-tensorial construction but instead it follows from the
energy-momentum tensor itself. This supports the well-defined
proposal developed by Cooperstock [42] and verified by many
authors [29-33,43]. In GR, many energy-momentum expressions
(reference frame dependent pseudo-tensors) have been proposed.
There is no consensus as to which is the best. Hamiltonian's
principle helps to solve this enigma. Each expression has a
geometrically and physically clear significance associated with
the boundary conditions.

\vspace{2cm}
\begin{description}
\item  {\bf Acknowledgment}
\end{description}

We would like to thank the unknown referee for his useful
comments.

\vspace{2cm}

{\bf \large References}

\begin{description}

\item{[1]} M\"{o}ller, C.: {\it The Theory of Relativity} (Oxford
University Press, 1957).

\item{[2]} Trautman, A.: Gravitation: {\it An Introduction to Current
Research} ed. Witten, L. (Wiley, New York, 1962)169.

\item{[3]} Landau, L.D. and  Lifshitz, E.M.: {\it The Classical Theory of Fields}
(Addison-Wesley Press, 1987).

\item{[4]} Tolman, R.C.: {\it Relativity, Thermodynamics and Cosmology} (Oxford
Univ. Press, London, 1934).

\item{[5]} Papapetrou, A.: {\it Proc. R. Irish Acad.} {\bf A52}(1948)11.

\item{[6]} M\"{o}ller, C.: Ann. Phys. (NY) {\bf 4}(1958)347.

\item{[7]} M\"{o}ller, C.: Ann. Phys. (NY) \textbf{12}(1961)118.

\item{[8]} Weinberg, I.: {\it Gravitation and Cosmology: Principle and Applications of
 General Theory of Relativity} (Wiley, New York, 1972)165.

\item{[9]} Bergmann, P.G. and Thompson, R.: Phys. Rev. {\bf D89}(1953)400.

\item{[10]} Misner, C.W., Thorne, K.S. and Wheeler, J.A.: {\it
Gravitation} (W. H. Freeman and Co., NY 1973)603.

\item{[11]} Cooperstock, F.I. and Sarracino, R.S.: J. Phys. A: Math. Gen.
\textbf{11}(1978)877.

\item{[12]} Bondi, H.: {\it Proc. R. Soc. London } \textbf{A427}(1990)249.

\item{[13]} Penrose, R.: \textit{Proc. Roy. Soc.} London
\textbf{A388}(1982)457;

GR 10 \textit{Conference} eds. Bertotti, B., de Felice, F. and
Pascolini, A. Padova \textbf{1}(1983)607.

\item{[14]} Brown, J.D. and York, Jr. J.W.: Phys. Rev. {\bf
D47}(1993)1407.

\item{[15]} Hayward, S.A.: Phys. Rev. {\bf D49}(1994)831.

\item{[16]} Bergqvist, G.: Class. Quantum Gravit. {\bf 9}(1992)1753.

\item{[17]} Bernstein, D.H. and Tod, K.P.: Phys. Rev. {\bf
D49}(1994)2808.

\item{[18]}  Virbhadra, K.S.: Phys. Rev. {\bf D41}(1990)1086; {\bf
D42}(1990)1066; and references therein.

\item{[19]} Virbhadra, K.S.: Phys. Rev. {\bf D60}(1999)104041.

\item{[20]} Virbhadra, K.S.: Phys. Rev. {\bf D42}(1990)2919.

\item{[21]} Virbhadra, K.S. and Parikh, J.C.: Phys. Lett. {\bf B317}(1993)312.

\item{[22]} Virbhadra, K.S. and Parikh, J.C.: Phys. Lett. {\bf B331}(1994)302.

\item{[23]} Rosen, N. and Virbhadra, K.S.: Gen. Relativ. Gravit.
{\bf 25}(1993)429.

\item{[24]} Aguirregabiria, J.M., Chamorro, A. and Virbhadra, K.S.:
Gen. Relativ. Gravit. {\bf 28}(1996)1393.

\item{[25]} Chang, C.C., Nester, J.M. and Chen, C.: Phys. Rev. Lett. {\bf
83}(1999)1897.

\item{[26]} Xulu, S.S.: Int. J. of Mod. Phys. {\bf A15}(2000)2979;
Mod. Phys. Lett. {\bf A15}(2000)1151 and reference therein.

\item{[27]} Xulu, S.S.: Astrophys. Space Sci. {\bf 283}(2003)23-32.

\item{[28]} Yang, I.C. and Radinschi, I.: Chin. J. of Phys. {\bf 41}(2003)4;
{\bf 42}(2004)1.

\item{[29]} Sharif, M.: Int. J. of Mod. Phys. {\bf A17}(2002)1175;

\item{[30]} Sharif, M.: Int. J. of Mod. Phys. {\bf A18}(2003)4361;
Errata {\bf A19}(2004)1495.

\item{[31]} Sharif, M.: Int. J. of Mod. Phys. {\bf D13}(2004)1019.

\item{[32]} Ragab, M. Gad: gr-qc/0401039.

\item{[33]} Patashnick, Owen: gr-qc/0408086.

\item{[34]} Chen, Chiang-Mei and Nester, James M.:
Class. Quan. Grav. {\bf 16}(1999)1279.

\item{[35]} Kovacs, D.: Gen. Relativ. Gravit. {\bf 17}(1985)927.

\item{[36]} Novotny, J.: Gen. Relativ. Gravit. {\bf 19}(1987)1043.

\item{[37]} Lessner, G.: Gen. Relativ. Gravit. {\bf 28}(1996)527.

\item{[38]} Stephani, H., Kramer, D., MacCallum, M.A.H., Hoenselaers, C. and
Hearlt, E.: {\it Exact Solutions of Einstein's field Equation}
(Cambridge University Press, 2003).

\item{[39]} Tupper, B.O.J.: Class. Quantum Gravit. {\bf 1}(1984)71-80;
{\bf 2}(1985)427-430.

\item{[40]} Senovilla, J.M.M.: Phys. Rev. Lett. {\bf 64}(1990)2219.

\item{[41]} Fernandez, L. and Gonzalez, L.M.: gr-qc/0402119.

\item{[42]} Cooperstock, F.I.: Annals Phys. \textbf{282}(2000)215;\\
Found. Phys. \textbf{22}(1992)1011; \textbf{33}(2003)1033.

\item{[43]} Bringly, Thomas: Mod. Phys. Lett. \textbf{A17}(2002)157;\\
Found. Phys. \textbf{22}(1992)1011; \textbf{33}(2003)1033.

\end{description}

\end{document}